\documentclass[12pt]{article}
\usepackage{scicite}
\usepackage{amsmath,amssymb,color,graphicx,dcolumn,bm}
\usepackage[utf8]{inputenc}
\usepackage{color,soulutf8}
\usepackage{authblk}
\newenvironment{sciabstract}{%
\begin{quote} \bf}
{\end{quote}}

\title{Observation of three-photon bound states in a quantum nonlinear medium}

\author[1]{Qi-Yu Liang}
\affil[1]{Department of Physics and Research Laboratory of Electronics, Massachusetts Institute of Technology, Cambridge, Massachusetts 02139, USA}
\author[2]{Aditya V. Venkatramani}
\affil[2]{Department of Physics, Harvard University, Cambridge, Massachusetts 02138, USA}
\author[1]{Sergio H. Cantu}
\author[1]{Travis L. Nicholson}
\author[3,4]{Michael J. Gullans}
\affil[3]{Department of Physics, Princeton University, Princeton, New Jersey 08544, USA}
\author[4]{Alexey V. Gorshkov}
\affil[4]{Joint Quantum Institute and Joint Center for Quantum Information and Computer Science,
National Institute of Standards and Technology and University of Maryland, College Park, Maryland 20742, USA}
\author[5]{Jeff D. Thompson}
\affil[5]{Department of Electrical Engineering, Princeton University, Princeton, NJ 08544, USA}
\author[6]{Cheng Chin}
\affil[6]{James Franck Institute, Enrico Fermi Institute and Department of Physics, University of Chicago, Chicago, IL 60637, USA
}
\author[2]{Mikhail D. Lukin}

\author[1]{Vladan Vuleti\'{c}}

\date{}

\begin{document}

\graphicspath{{./figures_v11/}}
\baselineskip24pt

\maketitle 

\begin{sciabstract}
Bound states of massive particles, such as nuclei, atoms or molecules, constitute the bulk of the visible world around us. In contrast, photons typically only interact weakly. We report the observation of traveling three-photon bound states in a quantum nonlinear medium where the interactions between photons are mediated by atomic Rydberg states. Photon correlation and conditional phase measurements reveal the distinct bunching and phase features associated with three-photon and two-photon bound states. Such photonic trimers and dimers possess shape-preserving wavefunctions that depend on the constituent photon number. The observed bunching and strongly nonlinear optical phase are quantitatively described by an effective field theory (EFT) of Rydberg-induced photon-photon interactions, consistent with the presence of a substantial effective three-body force between the photons. These observations demonstrate the ability to realize and control strongly interacting quantum many-body states of light.
\end{sciabstract}

Bound states of light quanta have been proposed to exist in specifically engineered media with strong optical nonlinearities \cite{deutsch1992diphotons, shen2007strongly, drummond1997optical, cheng1995optical,PhysRevA.92.033803}. Recently photonic dimers have  been observed experimentally \cite{firstenberg2013attractive}. Such bound states of photons can be viewed as quantum solitons \cite{Drummond1993,lai1989quantum}, which are shape-preserving wave-packets  enabled by the cancellation of nonlinear and dispersive effects. In contrast to classical solitons where the self-consistent shape varies smoothly  with total pulse energy, in a quantum soliton the optical nonlinearity is so strong that the wave packet shape depends on the constituent number of photons in a quantized manner \cite{Drummond1993,lai1989quantum}. The creation of quantum solitons not only represents an important step in fundamental studies of photonic quantum matter \cite{firstenberg2013attractive,chang2008crystallization,PhysRevA.91.033838}, but also may enable new applications in areas ranging from quantum communication to quantum metrology \cite{Li2016, RevModPhys.84.777}. 

We search for a photonic trimer using an ultracold atomic gas as a quantum nonlinear medium. This medium is experimentally realized by coupling photons to highly excited atomic Rydberg states by means of electromagnetically induced transparency (EIT). The resulting hybrid excitations of light and matter -- Rydberg polaritons -- inherit strong interactions from their Rydberg components, and can propagate with very low loss at slow group velocity $v_g$ \cite{friedler2005long,PhysRevLett.107.213601,gorshkov2011photon}. 
The nonlinearity arises when photons are within a Rydberg blockade radius $r_B$ of one another, where strong interactions between atoms in the Rydberg state \cite{labuhn2016tunable} shift the Rydberg level out of the EIT resonance, blocking the excitation of more than one Rydberg atom within $r_B$. In the dissipative regime (on atomic resonance), the blockade results in photon loss and anti-bunching \cite{dudin2012strongly,peyronel2012quantum,PhysRevLett.110.103001}. In the dispersive, off-resonant regime, the index of refraction varies with the separation between photons, resulting in an attractive force \cite{firstenberg2013attractive}.
 
Our experimental setup \cite{thompson2017symmetry} (Fig 1A, B) consists of a weak quantum probe field at 780 nm coupled to the 100S$_{1/2}$ Rydberg state via a strong $479$ nm control field in the EIT configuration (see Fig. 1B). The interactions occur in a cloud of laser-cooled $^{87}$Rb atoms in a far-detuned optical dipole trap. The system is effectively one-dimensional for the photons, due to the blockade radius ($r_B=20$  ${\mu}\text{m}$) being large compared to the transverse extent of the probe beam waist ($w$=4.5 $\mu$m), but smaller than the atomic cloud along the propagation direction ($\sim$130 $\mu \text{m}$). Measurements are conducted at a peak optical depth per blockade radius $\text{OD}_\text{B} \simeq 5$. To suppress dissipative effects, we work at large detuning $\Delta\geq3\Gamma$ from atomic resonance ($\Gamma$ is the population decay rate of the $5P_{3/2}$ state, see Fig. 1B), and at a control laser Rabi frequency where the transmission through the medium is the same with and without EIT, but the phase differs appreciably (Fig. 1C). Consequently, the transmission hardly varies with probe photon rate (Fig. 1D top), while a strongly rate-dependent phase with a slope of 0.40(7) rad$\cdot\mu \text{s}$ is observed (Fig. 1D bottom).

The quantum dynamics of interacting photons are investigated by measuring the three-photon correlation function and phase. Because dispersion outside of the atomic medium is negligible, any amplitude and phase features formed inside the nonlinear medium are preserved outside, and can be detected in the form of photon number and phase correlations. The third-order photon correlation function has been measured previously in coupled atom-cavity and quantum dot-cavity systems, as well as in non-classical states of three photons such as the Greenberger-Horne-Zeilinger (GHZ) and `N00N' states \cite{RevModPhys.84.777}. In our approach, we split the light onto three single-photon counting modules. Furthermore,  by mixing a detuned local oscillator (LO) into the final beamsplitter, we can also perform a heterodyne measurement in one of the detection arms (Fig. 1A). To connect the observed correlations to the physics of interacting Rydberg polaritons, we consider a state containing up to three photons,
\begin{equation}
|\psi\rangle=|0\rangle+\int \! \!dt_1\,\psi_1(t_1)|t_1\rangle+\int \!\!dt_1  dt_2\, \psi_2(t_1,t_2)|t_1,t_2\rangle+\int\!\! dt_1 dt_2 dt_3\, \psi_3(t_1,t_2,t_3)|t_1,t_2,t_3\rangle,
\end{equation}
where $|t_1,\cdots,t_N\rangle=\frac{1}{N!}a^\dagger(t_1)\cdots a^\dagger(t_N)|0\rangle$, and $a^\dagger(t)$ is the photon creation operator of the time bin mode $t$. The correlation functions can be related to the wavefunctions as $g^{(2)}(t_1,t_2)=\frac{|\psi_2(t_1,t_2)|^2}{|\psi_1(t_1)|^2|\psi_1(t_2)|^2}$ and $g^{(3)}(t_1,t_2,t_3)=\frac{|\psi_3(t_1,t_2,t_3)|^2}{|\psi_1(t_1)|^2|\psi_1(t_2)|^2|\psi_1(t_3)|^2}$. We refer to the phase $\tilde{\phi}^{(N)}$ of the $N$-photon wavefunction $\psi_N$ as the $N$-photon phase, namely, $\tilde{\phi}^{(1)}(t_1)=\text{Arg}[\psi_1(t_1)]$, $\tilde{\phi}^{(2)}(t_1,t_2)=\text{Arg}[\psi_2(t_1,t_2)]$, and $\tilde{\phi}^{(3)}(t_1,t_2,t_3)=\text{Arg}[\psi_3(t_1,t_2,t_3)]$. The $N$-photon phase is obtained from the phase of the beat note signal on the third detector, conditioned on having observed $N$-1 photons in the other two detectors. The conditional phase relative to $N$ uncorrelated photons, i.e. the nonlinear part of the phase, is denoted as $\phi^{(N)}$ (Fig. 3).

The experimentally measured $g^{(3)}$ function (Fig. 2A, B) displays a clear bunching feature: the probability to detect three photons within a short time ($\lesssim$ 25 ns) of one another is six times larger than for non-interacting photons in a laser beam. The increase at $t_1=t_2=t_3$ is accompanied by a depletion region for photons arriving within $\sim0.7$ $\mu s$ of one another, particularly visible along the lines of two-photon correlations $t_i=t_j\neq t_k$ (Fig. 2A): This depletion region is caused by the inflow of probability current towards the center $t_1=t_2=t_3$.
Figure 2B compares the two-photon correlation function $g^{(2)}(t,t+|\tau|)$ to that for three photons of which two photons were detected in the same time bin, $g^{(3)}(t,t,t+|\tau|)$. The trimer feature is approximately a factor of 2 narrower than the dimer feature, showing that a photon is attracted more strongly to two other photons than to one. Figure 2C illustrates the binding of a third photon to two photons that are detected with a time separation $T$. If $T$ exceeds the dimer time scale $\tau_2$, then the third photon binds independently to either photon, while for $T<\tau_2$ the two peaks merge into a single, more tightly bound trimer. This is analogous to the binding of a particle to a double-well potential as the distance between the wells is varied, since the polaritons can be approximately described as interacting massive particles moving at finite group velocity \cite{firstenberg2013attractive}. 

The dispersive and distance-dependent photon-photon interaction also manifests itself in a large conditional phase shift that depends on the time interval $\tau$ between the detection of the conditioning photons (at times $t_1$ = $t_2$ = $t$) and the phase measurement on detector $D_3$ at time $t_3$. We observe a conditional phase shift $\phi^{(3)}(t,t,t+|\tau|)$ for the trimer near $\tau=0$ (Fig. 3A) that is significantly larger than the dimer phase shift $\phi^{(2)}(t,t+|\tau|)$ (Fig. 3B). This confirms the stronger interaction between a photon and a dimer compared to that between one photon and another. 

To understand these results quantitatively, we apply an effective field theory (EFT) \cite{PhysRevLett.117.113601} which describes the low-energy scattering of Rydberg polaritons. This EFT gives us a one-dimensional slow-light Hamiltonian density with a contact interaction.
\begin{equation}
\mathcal{H} = - \hat{\psi}^\dagger \left( i\hbar v_g \partial_z + \frac{\hbar^2}{2 m} \partial_z^2 \right) \hat{\psi} - \frac{\hbar^2}{m a} \hat{\psi}^{\dagger 2} \hat{\psi}^2 ,
\label{eqn:hamiltonian}
\end{equation}
where $v_g$ is the group velocity inside the medium, $m = -\hbar \Omega_c^2 / (8 \Delta v_g^2)$ is the effective photon mass, $a$ is the scattering length, $\Omega_c$ is the control laser Rabi frequency, and $\Delta$ is the one-photon detuning. For weak interactions, $a\approx 15.28 (\frac{1}{\text{OD}_\text{B}}\frac{\Delta}{\Gamma})^2r_B$ \cite{PhysRevA.90.053804,PhysRevLett.117.113601}. The contact model provides an acccurate description of the low-energy scattering whenever $a\gg r_B$, the microscopic range of the two-body potential. For our parameters, we find this is well satisfied as $a\gtrsim 10r_B$. $\hat{\psi}$ is a quantum field annihilation operator, which corresponds to a photon outside the medium and a Rydberg polariton inside. Note that for our blue-detuned probe, the effective mass is negative and the interaction is repulsive. This situation maps onto a system with a positive mass and attractive interaction. The transverse mass is substantially heavier than $m$ \cite{PhysRevLett.101.163601}, effectively freezing out the transverse degrees of freedom over the timescale of the experiment. The bound states can be determined from the exact solution of this model for finite particle numbers \cite{lieb1963exact,mcguire1964study}, resulting in the correlation functions $g^{(3)}(t_1,t_2,t_3) \propto e^{-\frac{\left| t_1 - t_2 \right| }{ a/(2v_g)}} e^{-\frac{\left| t_2 - t_3 \right| }{ a/(2v_g)}} e^{-\frac{\left| t_1 - t_3 \right| }{ a/(2v_g)}}$ and $g^{(2)}(t_1,t_2) \propto e^{-\frac{\left| t_1 - t_2 \right| }{ a/(2v_g)}}$. 


In the case $t_1 = t_2 = t$, we find that $g^{(3)}(t,t,t+|\tau|)\propto e^{-2 \frac{\left| \tau \right| }{ a/(2\text{v}_g)}}$, implying  that the width of three-photon wave-packet (corresponding to $g^{(3)}$) is half that of $g^{(2)}$ for the same experimental conditions, in good agreement with experimental observations.  We calculate $a/(2v_g) = 0.32$ $\mu \text{s}$ for our measured experimental parameters \cite{si} and find it to be consistent with data (Fig. 2B, dashed lines). Following the quantum quench at the entry of the medium, the initial state is decomposed into the bound state and the continuum of scattering states \cite{firstenberg2013attractive}. Near $\tau= 0$, the scattering states dephase with each other, while the bound state propagates without distortion \cite{si}. This leads to a small contribution of scattering states in this region, with the bound state dominating the $g^{(3)}$ function. The observed value of $g^{(3)}(0)$ is not universal, as it is affected by the contributions from long-wavelength scattering states and nonlinear losses in the system and, therefore, depends on the atomic density profile of the medium. The dimer and trimer binding energies can be estimated as $E_2 = -\frac{\hbar^2}{ma^2}$=$h\times$0.2 MHz and $E_3 = 4 E_2$ respectively. This binding energy is $\sim 10^{10}$ times smaller than in diatomic molecules such as NaCl and H$_2$, but is comparable to Feshbach \cite{RevModPhys.82.1225} and Efimov \cite{braaten2007efimov} bound states of atoms with similar mass $m$ and scattering length $a$. To further characterize the three-photon bound state, it is instructive to consider the phase ratio  $\phi^{(3)}/\phi^{(2)}$ . For the bound-state contribution to the conditional phase $\phi^{(3)}(t,t,t)$ ($\phi^{(2)}(t,t)$), the Hamiltonian of Eq.\ref{eqn:hamiltonian} predicts a phase that equals the trimer binding energy times the propagation time in the medium. Thus from the bound state contributions, one would expect a ratio $\phi^{(3)}/\phi^{(2)}=4$, independent of the atom-light detuning $\Delta$. While the observed ratio (Fig. 4B) is approximately constant, it is smaller than 4. 



The observed deviation is likely due to the  two contributions of comparable magnitude. One correction arises from the scattering states, or equivalently, from the fact that our Rydberg medium 
($\sim$130 ${\mu}\text{m}$) is comparable in size to the two-photon bound state ($\sim$280 $\mu \text{m}$).
For a medium that is short compared to the bound state, one expects the ratio to be 3, consistent with a dispersionless Kerr medium \cite{bienias2016quantum}. The other, more fundamental correction, may be due to a contribution that does not arise from pairwise interactions, effectively representing a three-photon force. Specifically, when all three photons are within one blockade radius of one another, there can be only one Rydberg excitation and the potential cannot exceed the value corresponding to that of two photons \cite{PhysRevLett.117.113601,jachymski2016three}. This saturation effect manifests itself as a short-range repulsive effective three-photon force which, according to our theoretical analysis \cite{si}, results in a reduction of $\phi^{(3)}/\phi^{(2)}$ below 3. The corresponding correction to the bound state is smaller in the weakly interacting regime relevant to these experiments \cite{jachymski2016three}. This explains why the effective three-photon force has a relatively weak effect on the bunching of $g^{(3)}(|\tau|<0.2$  $\mu\text{s})$, which is dominated by the bound state. Note that both the scaling arguments and numerical evidence indicate that the effective three-photon force contributes to the three-body scattering amplitudes more strongly than two-body finite range effects in this regime \cite{PhysRevLett.117.113601}.  



To quantitatively understand these effects, the EFT is modified to include the estimated effective three-photon force \cite{si}. Using the modified EFT, we compare the results with and without the repulsive effective three-photon force, while also taking into account the effects due to finite medium (Fig. 4B). Including this three-photon saturation force allows the phase ratio $\phi^{(3)}/\phi^{(2)}$ to go below 3, in a reasonable agreement with the experimental observations. For fully saturated interactions between the polaritons, the interaction potential does not increase with photon number, and the phase ratio should approach 2.



The observation of the three-photon bound state, which can be viewed as photonic solitons in the quantum regime \cite{Drummond1993,lai1989quantum}, can be extended along several different directions. First, increasing the length of the medium at constant atomic density would remove the effect of the scattering states through destructive quantum interference to larger $\tau$ and retain only the solitonic bound-state component. Additionally, the strong observed rate dependence of $\phi^{(3)}$ may indicate that larger photonic molecules and photonic clusters could be observed with improved detection efficiency and data acquisition rate. Furthermore, using an elliptical or larger round probe beam and carefully engineering the mass along different directions, the system can be extended to two and three dimensions, possibly permitting the observation of photonic Efimov states \cite{gullans2017efimov, kraemer2006evidence}. Finally, our medium only supports one two/three-photon bound state, corresponding to a nonlinear phase less than $\pi$. A threefold increase in the atomic density would render the interaction potential sufficiently deep for a second bound state to appear near zero energy, which should result in resonant photon-photon scattering and a tunable scattering length \cite{PhysRevA.90.053804}. The presence of large effective N-body forces in this system opens intriguing possibilities to study exotic many-body phases of light and matter, including self-organization in open quantum systems \cite{PhysRevLett.118.133401, Schausz2012}, and quantum materials that cannot be realized with conventional systems.

\clearpage
\begin{figure}[htbp]
\begin{center}
\includegraphics[width=.9\textwidth]{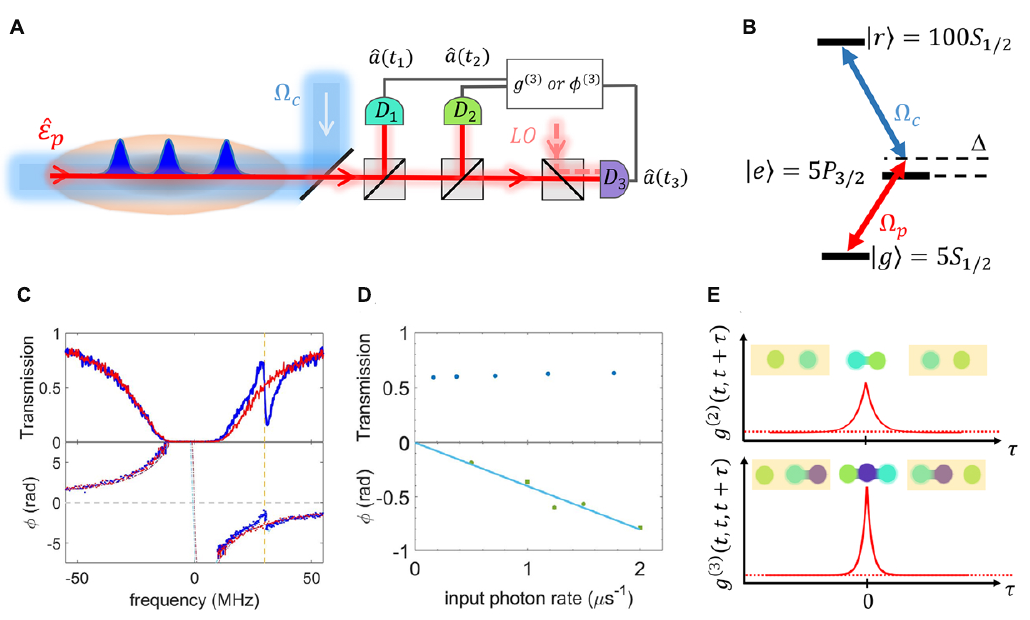}
\caption{\textbf{Qualitative descriptions of the experiment.} \textbf{A,B,} Setup and atomic level scheme. The atoms are optically pumped into the hyperfine ($F$) and magnetic ($m_F$) sublevel $|g\rangle=|5S_{1/2},F=2,m_F=2\rangle$. The weak coherent probe light is coupled to the Rydberg state $|r\rangle=|100S_{1/2}, m_J=1/2\rangle$, via an intermediate state $|e\rangle=|5P_{3/2}, F=3, m_F=3\rangle$, with linewidth $\Gamma/2\pi=6.1$ MHz, by means of a counter-propagating control field that is detuned by $\Delta$ below the resonance frequency of the upper transition, $|e\rangle\rightarrow|r\rangle$. Strong interactions between probe photons are detected via photon correlations of the transmitted light, which is split onto three single-photon detectors with equal intensities. To perform phase measurements, a local oscillator is mixed into detector $D_3$. \textbf{C}, Transmission (top) and phase $\phi$ (bottom) as a function of probe frequency measured at a low (0.5 ${\mu}\text{s}^{-1}$) input photon rate. $\phi$ is measured without conditioning on the detection of other photons. The control laser is set at $\Delta/ 2\pi = 30 $ MHz below the $|e\rangle\rightarrow|r\rangle$ transition with Rabi frequency $\Omega_c /2\pi = 10$ MHz. The blue and red data are from measurements with and without control beam, respectively. The blue and red dashed lines in the bottom graph are theoretical expectations. The vertical yellow dashed line marks EIT resonance. \textbf{D}, Rate dependence of transmission (top) and unconditional phase (bottom) on two-photon resonance $|g\rangle \rightarrow |r\rangle$, with a one-photon detuning of $\Delta/ 2\pi = 30$ MHz, and control Rabi frequency $\Omega_c /2\pi = 10$ MHz. While the transmission is rate-independent, the phase is strongly rate dependent (slope is 0.4 rad$\cdot\mu \text{s}$). \textbf{E}. Schematic correlation functions for two (top) and three (bottom) photons as a function of their time separation $\tau$. The attractive interaction leads to photon bunching, with three photons being more tightly bound together than two photons.
}
\label{default}
\end{center}
\end{figure}

\begin{figure}[htbp]
\begin{center}
\includegraphics[width=.78\textwidth]{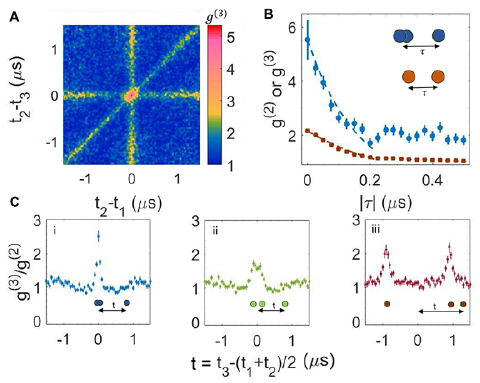}
\caption{ \textbf{Photon correlation functions with tighter bunching due to the three-photon bound state.} Photon correlation functions measured at one-photon detuning $\Delta/ 2\pi=30$ MHz, control Rabi frequency $\Omega_c/2\pi=10$ MHz, input photon rate of 1 ${\mu}s^{-1}$ and on EIT resonance. \textbf{A}, Two-dimensional representation of three-photon correlation function $g^{(3)}(t_1,t_2,t_3)$, with $t_i$ being the photon detection time at detector $D_i$. Three-photon bunching corresponds to the central region, two-photon bunching to the stripes. \textbf{B}, $g^{(3)}(t,t,t+|\tau|)$ (blue data points) and $g^{(2)}(t,t+|\tau|)$ (brown data points), with the decay constants calculated from the exact solution for the bound states $\tau^c_3=0.16$ $\mu s$ and $\tau^c_2=0.32$ ${\mu} \text{s}$ respectively (dashed lines). The calculated exponential decay is scaled to match the initial point of the measured intensity correlation functions. The approximately twice smaller decay length of the three-photon correlation function shows that a photon is more strongly bound to two photons than to one. The fitted exponential decay constants with zero offset for $g^{(3)}$ and $g^{(2)}$ are $\tau_3=0.14(2)$ $\mu s$ and $\tau_2=0.31(6)$ ${\mu} \text{s}$, respectively (not shown), in agreement with the calculated values.  \textbf{C}, Three representative plots of $g^{(3)}(t_1, t_2, t_3)/g^{(2)}(t_1,t_2)$ for fixed $T\equiv|t_1-t_2|=0$ $\mu s$ (i), $T=0.2$ $\mu \text{s}$ (ii), and $T=1.8$ $\mu \text{s}$ (iii), within a $50\text{ ns}$ window. As we condition on the two photons being further and further away, the sharply decaying $g^{(3)}$ function transitions to a slower decaying $g^{(2)}$ function. For intermediate time separations (ii), there is interference between all states including the dimer and trimer. All permutations of the detectors are used to generate the data in B,C. Error bars in figure indicate one standard deviation (s.d). Error bars in the fitted exponential decay constants indicate one s.d of the fit.
}
\label{default}
\end{center}
\end{figure}

\begin{figure}[htbp]
\begin{center}
\includegraphics[width=.9\textwidth]{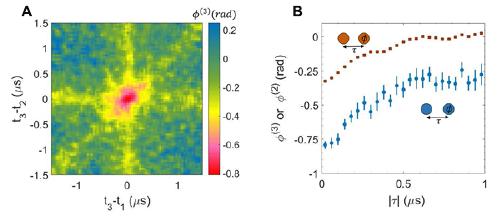}
\caption{ \textbf{Larger nonlinear phase for three photons.} Nonlinear phase measured under identical conditions as the data in Fig. 2. \textbf{A}, Conditional phase $\phi^{(3)}(t_1,t_2,t_3)$, where $t_1$ and $t_2$ correspond to photon detection events at detectors $D_1$, $D_2$, and a heterodyne measurement is performed on detector $D_3$ at time $t_3$. \textbf{B}, Diagonal cut $\phi^{(3)}(t,t,t+|\tau|)$ (blue), with the two conditioning probe photons within $40\text{ ns}$ of each other, and $\phi^{(2)}(t,t+|\tau|)$ (brown), showing a larger phase when conditioning on two other near-simultaneous photons ($\phi^{(3)}$) than on one near-simultaneous photon ($\phi^{(2)}$). $\phi^{(N)}$ is referenced to itw own average value when all the N photons are too far apart from each other to be correlated. Specifically, $\phi^{(2)}(t_1,t_2)\equiv\tilde{\phi}^{(2)}(t_1,t_2)-(\tilde{\phi}^{(1)}(t_1)+\tilde{\phi}^{(1)}(t_2))\xrightarrow[]{|t_1-t_2|\rightarrow\infty}0$, and $\phi^{(3)}(t_1,t_2,t_3)\equiv\tilde{\phi}^{(3)}(t_1,t_2,t_3)-(\tilde{\phi}^{(1)}(t_1)+\tilde{\phi}^{(1)}(t_2)+\tilde{\phi}^{(1)}(t_3))\xrightarrow[]{|t_i-t_j|\rightarrow\infty, \forall i \neq j}0$.  $\phi^{(3)}$ at large $|\tau|$ asymptotically goes to $\phi^{(2)}(t,t)$, because $\phi^{(3)}(t,t,t+|\tau|) \xrightarrow[]{|\tau|\rightarrow\infty} \tilde{\phi}^{(2)}(t,t)+\tilde{\phi}^{(1)}(t+|\tau|)-(\tilde{\phi}^{(1)}(t)+\tilde{\phi}^{(1)}(t)+\tilde{\phi}^{(1)}(t+|\tau|))= \phi^{(2)}(t,t)$. Error bars indicate one s.d.
}
\label{default}
\end{center}
\end{figure}

\begin{figure}[htbp]
\begin{center}
\includegraphics[width=.9\textwidth]{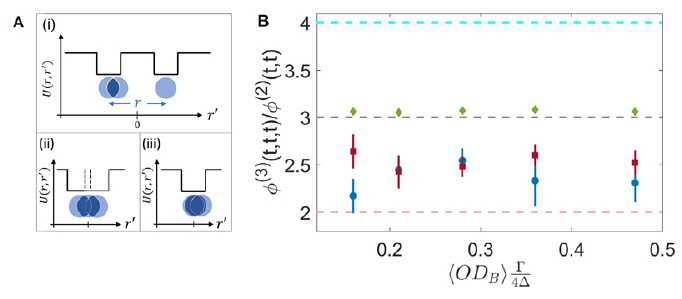}
\caption{ \textbf{Comparison of the phase ratio with the EFT predictions.} \textbf{A} illustrates the potential (solid black and gray lines) the third photon, at position $r'$, experiences due to the other two photons, at positions $\pm r/2$. (i) When the two photons are separated by more than twice the blockade radius ($r>2r_B$), each of them creates its own square potential with a width of $2r_B$; (ii) When the two photons overlap ($r_B<r<2r_B$), the potential is partially saturated; (iii) When the two photons are within one blockade radius ($r<r_B$), since there can be at most one Rydberg excitation within $r_B$, the potential is not deeper than that created by one photon. Therefore, we overestimate the attractive potential by considering pairwise interaction only, and a repulsive effective three-photon force is required to correctly take into account the saturation of the Rydberg blockade. \textbf{B}, Measured phase ratio $\phi^{(3)}(t,t,t)/\phi^{(2)}(t,t)$ (blue) and the EFT predictions (with the effective three-photon force in brown; without in green) as a function of $\langle \text{OD}_\text{B}\rangle\frac{\Gamma}{4\Delta}$, where $\langle\rangle$ refers to the average over the Gaussian profile of the atomic density. The quantity $\langle \text{OD}_\text{B}\rangle\frac{\Gamma}{4\Delta}$ is a quantitative measure of the interaction strength in this system. The control Rabi frequency $\Omega_c/ 2\pi=\{$22,18,10,10,8$\}$ MHz for $\Delta/ 2\pi=\{$54,42,30,24,18$\}$ MHz is chosen such that the transmission is insensitive to the input photon rate (Fig. 1C). We also change the input photon rate $\{0.7,1,1,1.3,2.5\}$ photons$/\mu \text{s}$ to achieve similar data acquisition rates since the losses are larger at smaller detunings. For a fully saturated medium, one expects $\phi^{(3)}/\phi^{(2)}=2$, as indicated by the pink dashed line; for bound states in a long medium and no effective three-photon force, one expects $\phi^{(3)}/\phi^{(2)}=4$, as indicated by the cyan dashed line (see text). EFT results are calculated with parameters from independent measurements, and the two-photon detuning from the EIT resonance is the only parameter varied within the experimental uncertainty to fit the two-photon phase. Error bars in the EFT with the effective three-photon force arise from the variations with the choice of matching conditions for the three-body scattering amplitudes \cite{si}. Error bars in the experimental data indicate one s.d. 
}
\label{default}
\end{center}
\end{figure}

\paragraph{Acknowledgments:} We thank O. Firstenberg for early stages of this work and S. Choi for discussions. MJG and AVG thank H. P. B{\"u}chler for many insightful discussions and comments on the theoretical analysis. This work has been supported by NSF, NSF CUA, ARO, AFOSR, ARO MURI and Bush Fellowship. AVG and MJG acknowledge support by ARL CDQI, NSF QIS, and NSF PFC at JQI. CC acknowledges funding support from NSF grant PHY-1511696 and Alexander von Humboldt foundation. Data are available upon request.

\section*{Supplementary Materials}

Materials and Methods

\noindent Supplementary Text

\noindent Figs. S1 to S4

\noindent Tables S1 to S2

\noindent References (\textit{35}-\textit{40})

\clearpage

\bibliography{three_photon_v11_rev5.bbl}

\begin{thebibliography}{10}

\bibitem{deutsch1992diphotons}
I.~H. Deutsch, R.~Y. Chiao, J.~C. Garrison, Diphotons in a nonlinear
  fabry-p{\'e}rot resonator: bound states of interacting photons in an optical
  ‘‘quantum wire’’, {\it Phys. Rev. Lett.\/} {\bf 69}, 3627 (1992).

\bibitem{shen2007strongly}
J.-T. Shen, S.~Fan, Strongly correlated two-photon transport in a
  one-dimensional waveguide coupled to a two-level system, {\it Phys. Rev.
  Lett.\/} {\bf 98}, 153003 (2007).

\bibitem{drummond1997optical}
P.~Drummond, H.~He, Optical mesons, {\it Phys. Rev. A\/} {\bf 56}, R1107
  (1997).

\bibitem{cheng1995optical}
Z.~Cheng, G.~Kurizki, Optical “multiexcitons”: quantum gap solitons in
  nonlinear bragg reflectors, {\it Phys. Rev. Lett.\/} {\bf 75}, 3430 (1995).

\bibitem{PhysRevA.92.033803}
Y.~Shen, J.-T. Shen, Photonic-fock-state scattering in a waveguide-qed system
  and their correlation functions, {\it Phys. Rev. A\/} {\bf 92}, 033803
  (2015).

\bibitem{firstenberg2013attractive}
O.~Firstenberg, {\it et~al.\/}, Attractive photons in a quantum nonlinear
  medium, {\it Nature\/} {\bf 502}, 71 (2013).

\bibitem{Drummond1993}
P.~D. Drummond, R.~M. Shelby, S.~R. Friberg, Y.~Yamamoto, {Quantum solitons in
  optical fibres}, {\it Nature\/} {\bf 365}, 307 (1993).

\bibitem{lai1989quantum}
Y.~Lai, H.~Haus, Quantum theory of solitons in optical fibers. ii. exact
  solution, {\it Phys. Rev. A\/} {\bf 40}, 854 (1989).

\bibitem{chang2008crystallization}
D.~Chang, {\it et~al.\/}, Crystallization of strongly interacting photons in a
  nonlinear optical fibre, {\it Nature Phys.\/} {\bf 4}, 884 (2008).

\bibitem{PhysRevA.91.033838}
M.~F. Maghrebi, {\it et~al.\/}, Fractional quantum hall states of rydberg
  polaritons, {\it Phys. Rev. A\/} {\bf 91}, 033838 (2015).

\bibitem{Li2016}
L.~Li, A.~Kuzmich, Quantum memory with strong and controllable rydberg-level
  interactions, {\it Nature Comm.\/} {\bf 7} (2016).

\bibitem{RevModPhys.84.777}
J.-W. Pan, {\it et~al.\/}, Multiphoton entanglement and interferometry, {\it
  Rev. Mod. Phys.\/} {\bf 84}, 777 (2012).

\bibitem{friedler2005long}
I.~Friedler, D.~Petrosyan, M.~Fleischhauer, G.~Kurizki, Long-range interactions
  and entanglement of slow single-photon pulses, {\it Phys. Rev. A\/} {\bf 72},
  043803 (2005).

\bibitem{PhysRevLett.107.213601}
D.~Petrosyan, J.~Otterbach, M.~Fleischhauer, Electromagnetically induced
  transparency with rydberg atoms, {\it Phys. Rev. Lett.\/} {\bf 107}, 213601
  (2011).

\bibitem{gorshkov2011photon}
A.~V. Gorshkov, J.~Otterbach, M.~Fleischhauer, T.~Pohl, M.~D. Lukin,
  Photon-photon interactions via rydberg blockade, {\it Phys. Rev. Lett.\/}
  {\bf 107}, 133602 (2011).

\bibitem{labuhn2016tunable}
H.~Labuhn, {\it et~al.\/}, Tunable two-dimensional arrays of single rydberg
  atoms for realizing quantum ising models, {\it Nature\/} {\bf 534}, 667
  (2016).

\bibitem{dudin2012strongly}
Y.~Dudin, A.~Kuzmich, Strongly interacting rydberg excitations of a cold atomic
  gas, {\it Science\/} {\bf 336}, 887 (2012).

\bibitem{peyronel2012quantum}
T.~Peyronel, {\it et~al.\/}, Quantum nonlinear optics with single photons
  enabled by strongly interacting atoms, {\it Nature\/} {\bf 488}, 57 (2012).

\bibitem{PhysRevLett.110.103001}
D.~Maxwell, {\it et~al.\/}, Storage and control of optical photons using
  rydberg polaritons, {\it Phys. Rev. Lett.\/} {\bf 110}, 103001 (2013).

\bibitem{thompson2017symmetry}
J.~D. Thompson, {\it et~al.\/}, Symmetry-protected collisions between strongly
  interacting photons, {\it Nature\/} {\bf 542}, 206 (2017).

\bibitem{PhysRevLett.117.113601}
M.~J. Gullans, {\it et~al.\/}, Effective field theory for rydberg polaritons,
  {\it Phys. Rev. Lett.\/} {\bf 117}, 113601 (2016).

\bibitem{PhysRevA.90.053804}
P.~Bienias, {\it et~al.\/}, {Scattering resonances and bound states for
  strongly interacting Rydberg polaritons}, {\it Phys. Rev. A\/} {\bf 90},
  53804 (2014).

\bibitem{PhysRevLett.101.163601}
M.~Fleischhauer, J.~Otterbach, R.~G. Unanyan, Bose-einstein condensation of
  stationary-light polaritons, {\it Phys. Rev. Lett.\/} {\bf 101}, 163601
  (2008).

\bibitem{lieb1963exact}
E.~H. Lieb, W.~Liniger, Exact analysis of an interacting bose gas. i. the
  general solution and the ground state, {\it Phys. Rev.\/} {\bf 130}, 1605
  (1963).

\bibitem{mcguire1964study}
J.~B. McGuire, Study of exactly soluble one-dimensional n-body problems, {\it
  Journal of Mathematical Physics\/} {\bf 5}, 622 (1964).

\bibitem{si}
Supplementary information is available at the Science Web site.

\bibitem{RevModPhys.82.1225}
C.~Chin, R.~Grimm, P.~Julienne, E.~Tiesinga, Feshbach resonances in ultracold
  gases, {\it Rev. Mod. Phys.\/} {\bf 82}, 1225 (2010).

\bibitem{braaten2007efimov}
E.~Braaten, H.-W. Hammer, Efimov physics in cold atoms, {\it Annals of
  Physics\/} {\bf 322}, 120 (2007).

\bibitem{bienias2016quantum}
P.~Bienias, H.~P. B{\"u}chler, Quantum theory of kerr nonlinearity with rydberg
  slow light polaritons, {\it New Journal of Physics\/} {\bf 18}, 123026
  (2016).

\bibitem{jachymski2016three}
K.~Jachymski, P.~Bienias, H.~P. B{\"u}chler, Three-body interaction of rydberg
  slow-light polaritons, {\it Phys. Rev. Lett.\/} {\bf 117}, 053601 (2016).

\bibitem{gullans2017efimov}
M.~Gullans, {\it et~al.\/}, Efimov states of strongly interacting photons, {\it
  arXiv preprint arXiv:1709.01955\/}  (2017).

\bibitem{kraemer2006evidence}
T.~Kraemer, {\it et~al.\/}, Evidence for efimov quantum states in an ultracold
  gas of caesium atoms, {\it Nature\/} {\bf 440}, 315 (2006).

\bibitem{PhysRevLett.118.133401}
N.~Thaicharoen, A.~Schwarzkopf, G.~Raithel, Control of spatial correlations
  between rydberg excitations using rotary echo, {\it Phys. Rev. Lett.\/} {\bf
  118}, 133401 (2017).

\bibitem{Schausz2012}
P.~Schausz, {\it et~al.\/}, Observation of spatially ordered structures in a
  two-dimensional rydberg gas, {\it Nature\/} {\bf 491}, 87 (2012).

\bibitem{carmichael1991quantum}
H.~Carmichael, R.~Brecha, P.~Rice, Quantum interference and collapse of the
  wavefunction in cavity qed, {\it Optics communications\/} {\bf 82}, 73
  (1991).

\bibitem{adhikari1995perturbative}
S.~K. Adhikari, T.~Frederico, I.~Goldman, Perturbative renormalization in
  quantum few-body problems, {\it Phys. Rev. Lett.\/} {\bf 74}, 487 (1995).

\bibitem{braaten2006universality}
E.~Braaten, H.-W. Hammer, Universality in few-body systems with large
  scattering length, {\it Physics Reports\/} {\bf 428}, 259 (2006).

\bibitem{bedaque1999renormalization}
P.~F. Bedaque, H.-W. Hammer, U.~Van~Kolck, Renormalization of the three-body
  system with short-range interactions, {\it Phys. Rev. Lett.\/} {\bf 82}, 463
  (1999).

\bibitem{larre2015propagation}
P.-{\'E}. Larr{\'e}, I.~Carusotto, Propagation of a quantum fluid of light in a
  cavityless nonlinear optical medium: General theory and response to quantum
  quenches, {\it Phys. Rev. A\/} {\bf 92}, 043802 (2015).

\bibitem{gullans2013controlling}
M.~J. Gullans, Controlling atomic, solid-state and hybrid systems for quantum
  information processing, Ph.D. thesis, Harvard University (2013).

\end{thebibliography}

\nocite{carmichael1991quantum} 
\nocite{adhikari1995perturbative}
\nocite{braaten2006universality}
\nocite{bedaque1999renormalization}
\nocite{larre2015propagation}
\nocite{gullans2013controlling}

\pagestyle{empty}
{
\begin{figure*}
\vspace{-3.35cm}
\hspace*{-2.75cm} 
\includegraphics[page=1]{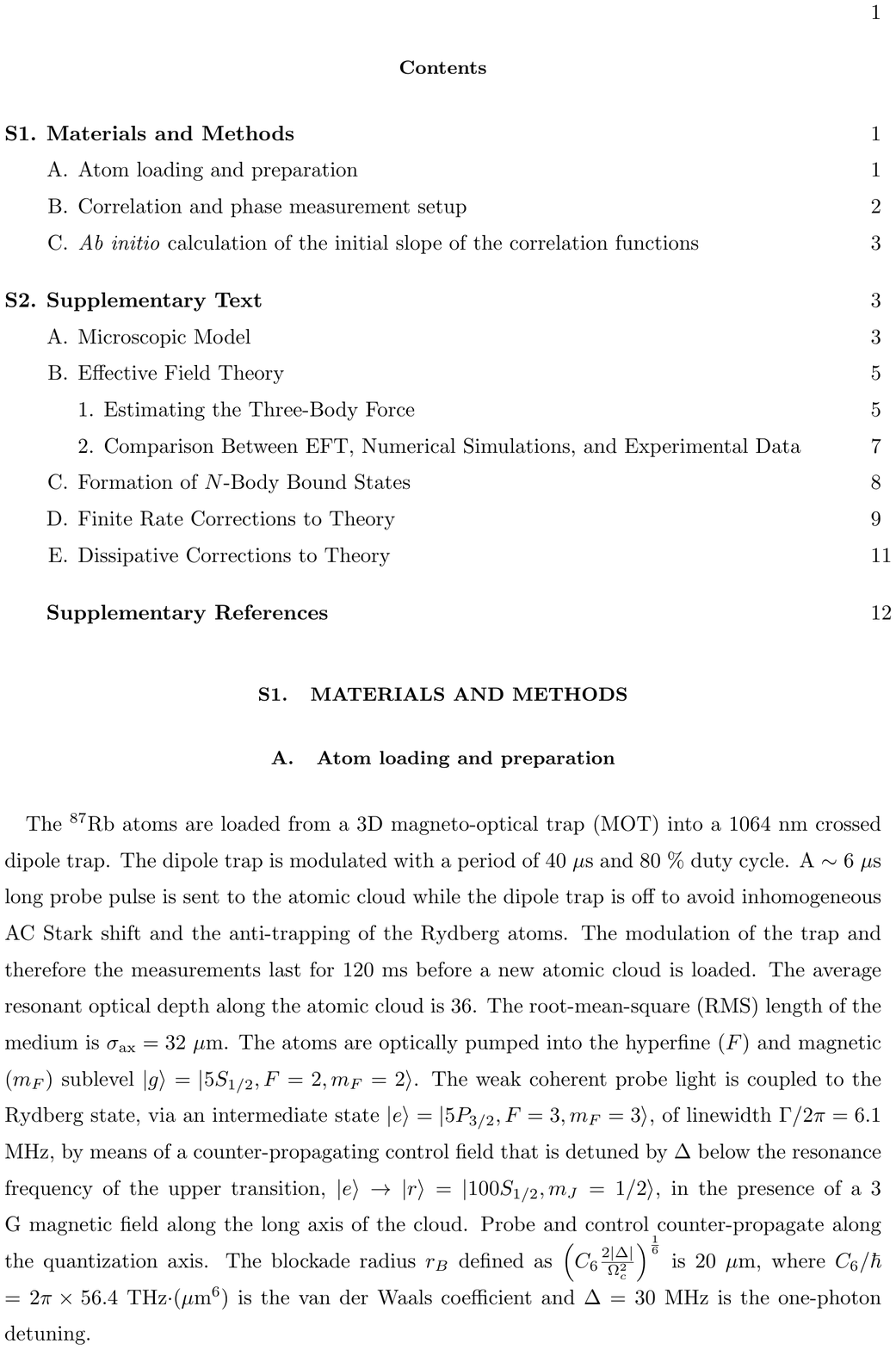}
\end{figure*}

\begin{figure*}
\vspace{-3.35cm}
\hspace*{-2.75cm} 
\includegraphics[page=2]{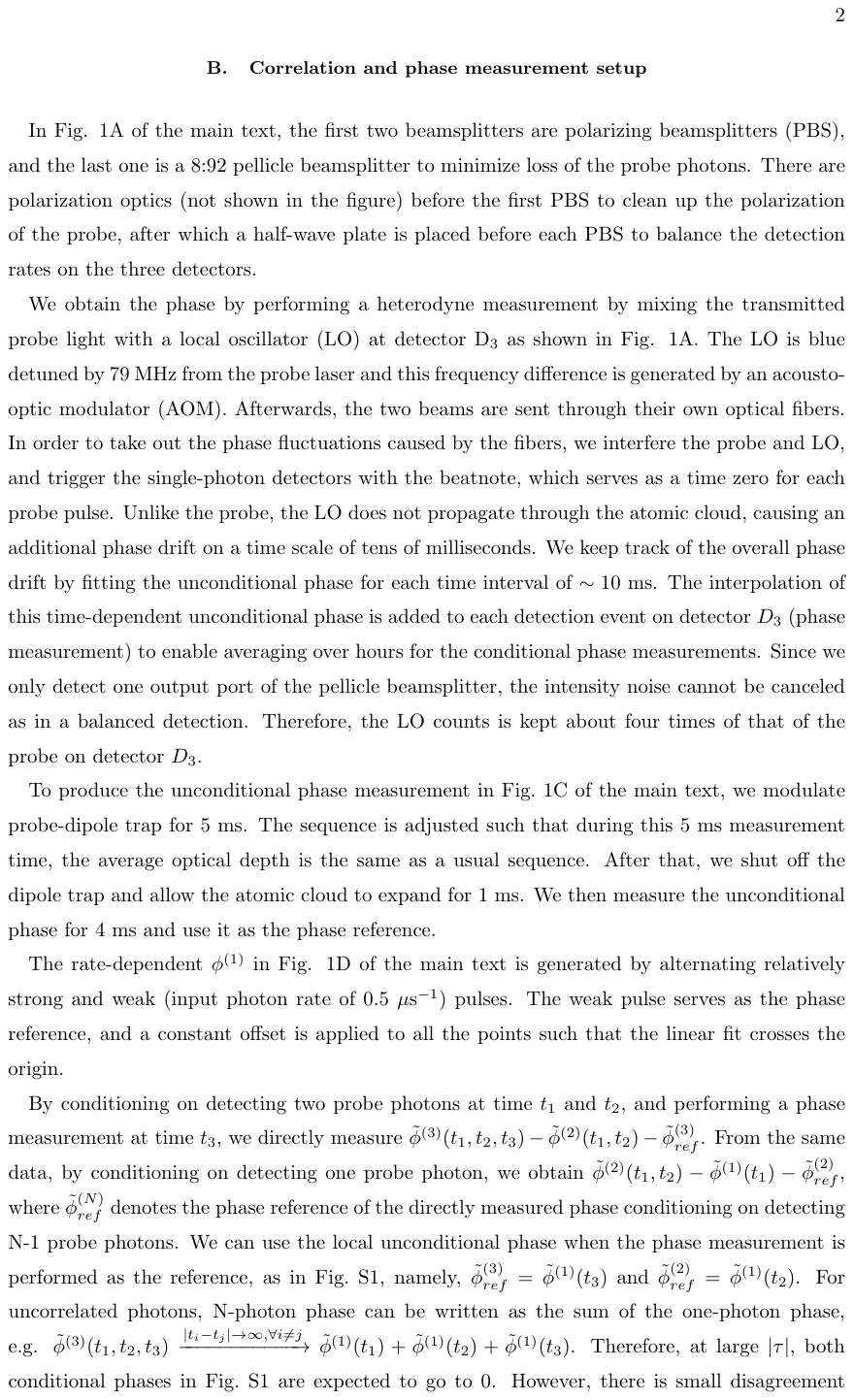}
\end{figure*}

\begin{figure*}
\vspace{-3.35cm}
\hspace*{-2.75cm} 
\includegraphics[page=3]{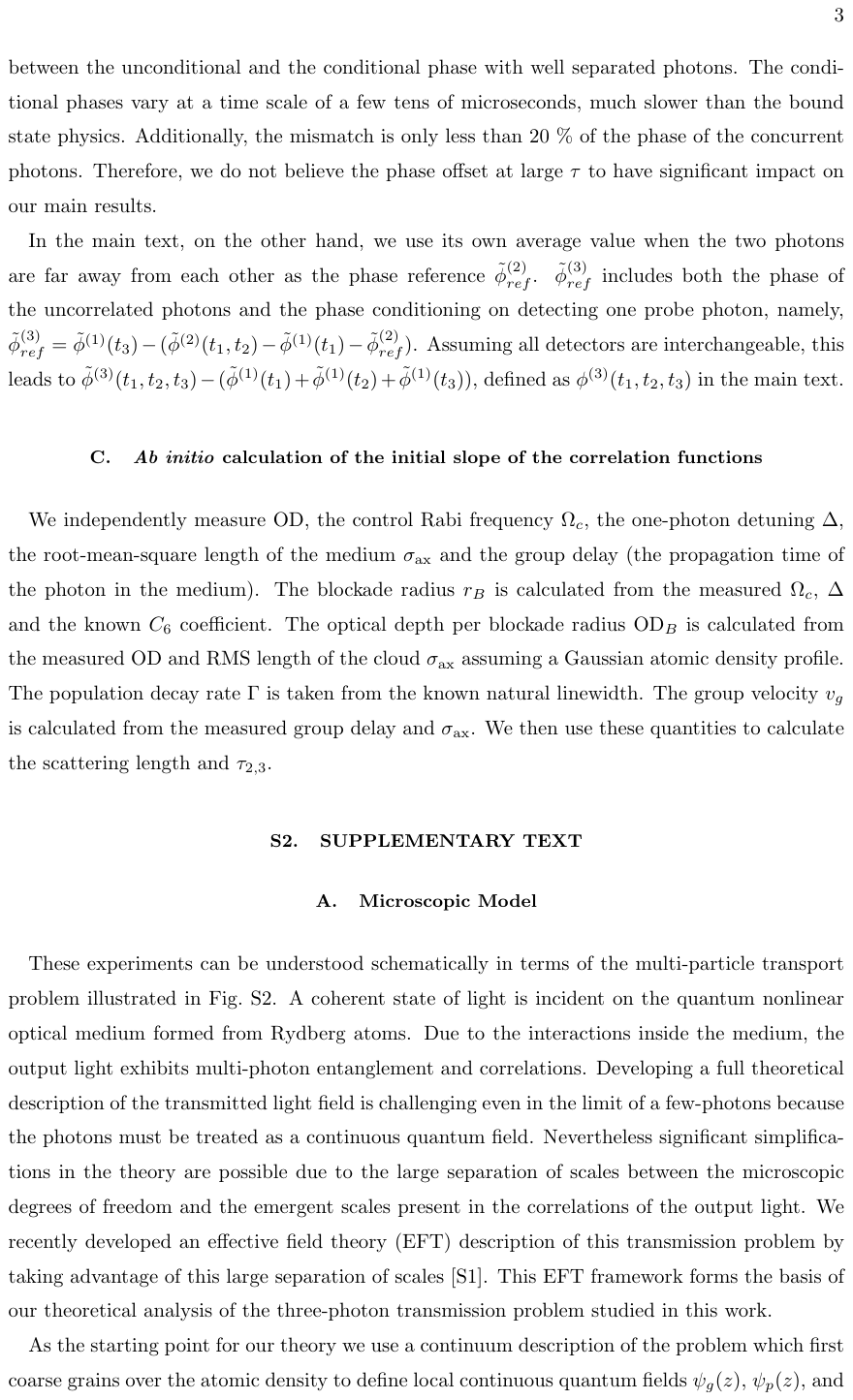}
\end{figure*}

\begin{figure*}
\vspace{-3.35cm}
\hspace*{-2.75cm} 
\includegraphics[page=4]{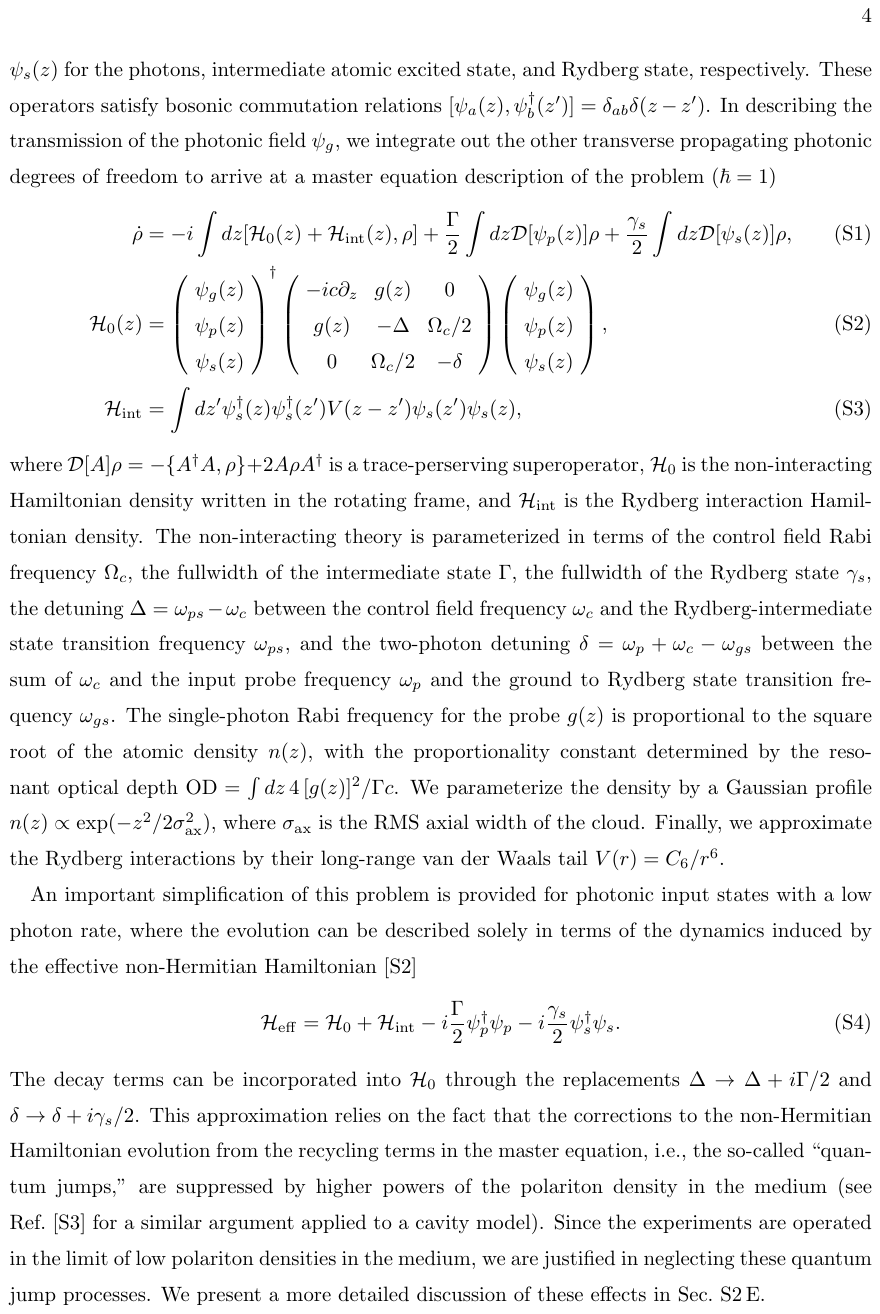}
\end{figure*}

\begin{figure*}
\vspace{-3.35cm}
\hspace*{-2.75cm} 
\includegraphics[page=5]{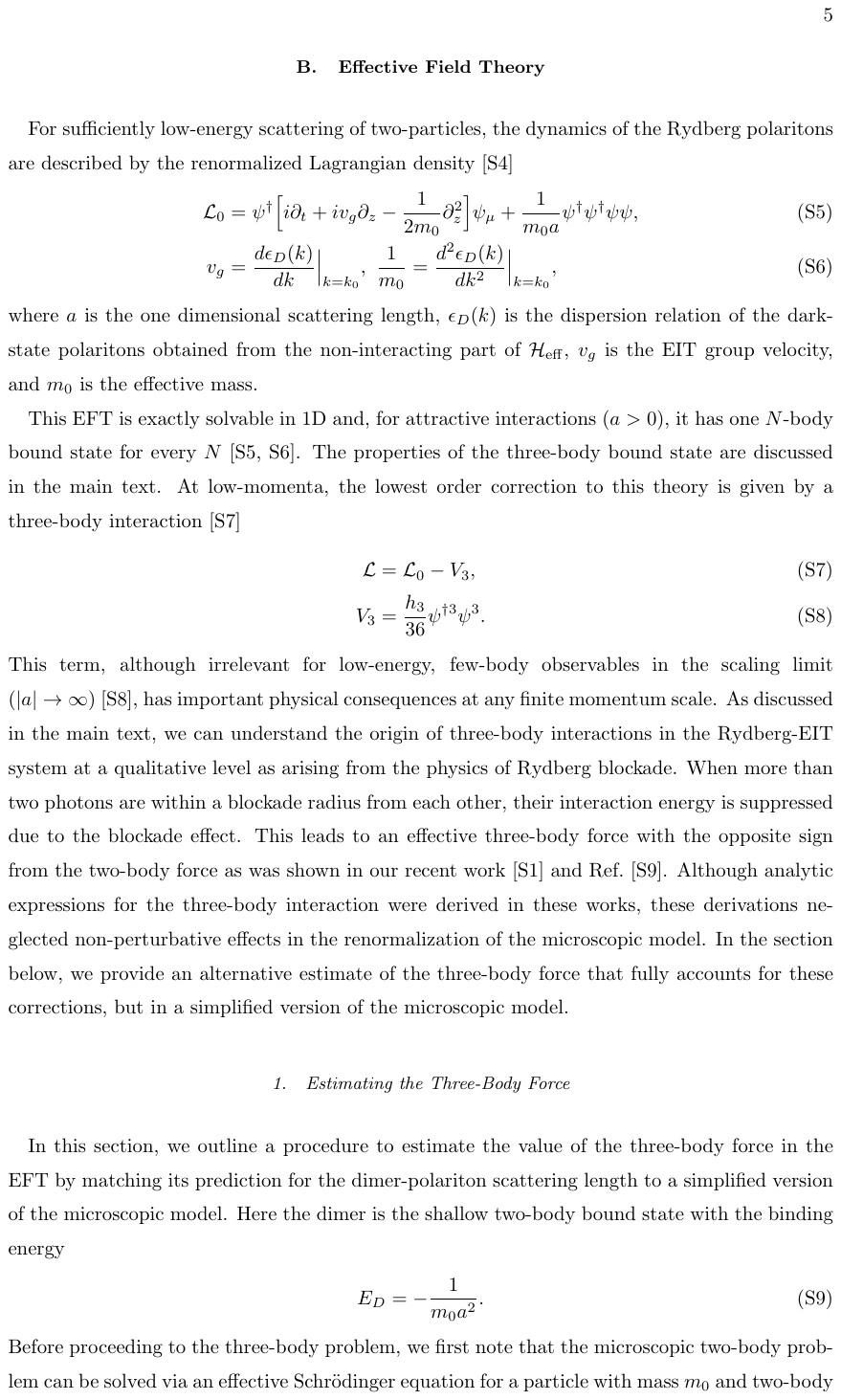}
\end{figure*}

\begin{figure*}
\vspace{-3.35cm}
\hspace*{-2.75cm} 
\includegraphics[page=6]{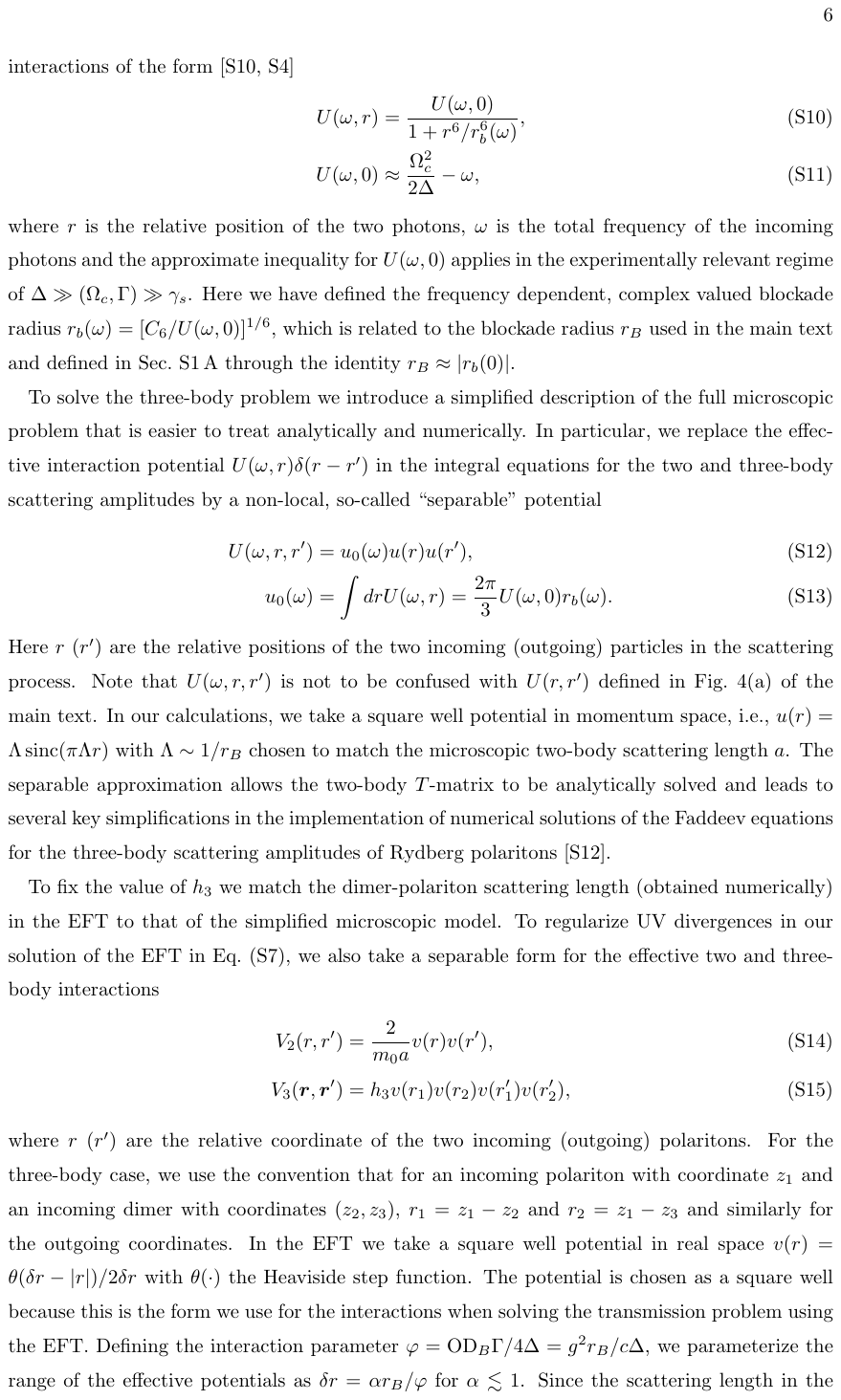}
\end{figure*}

\begin{figure*}
\vspace{-3.35cm}
\hspace*{-2.75cm} 
\includegraphics[page=7]{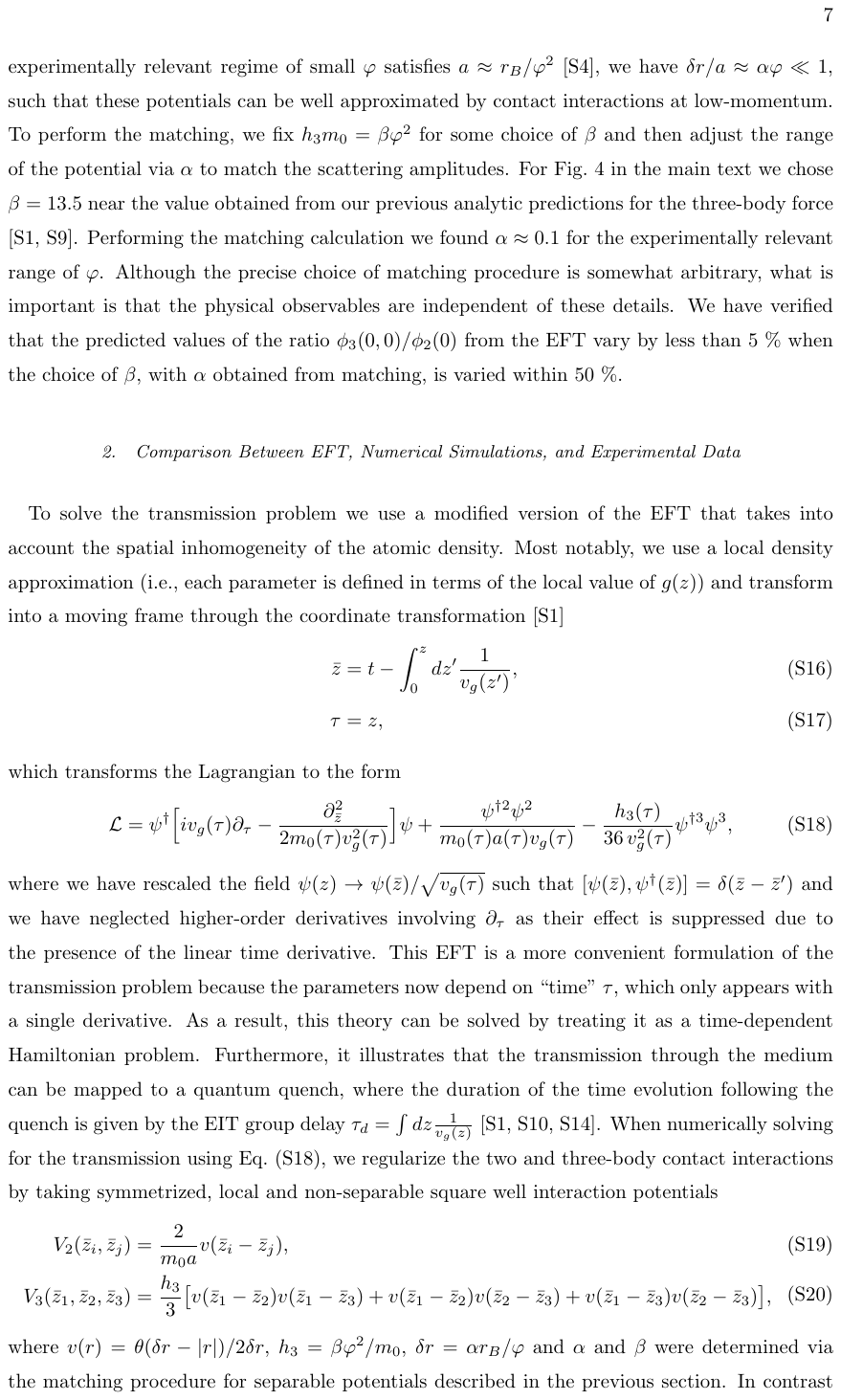}
\end{figure*}

\begin{figure*}
\vspace{-3.35cm}
\hspace*{-2.75cm} 
\includegraphics[page=8]{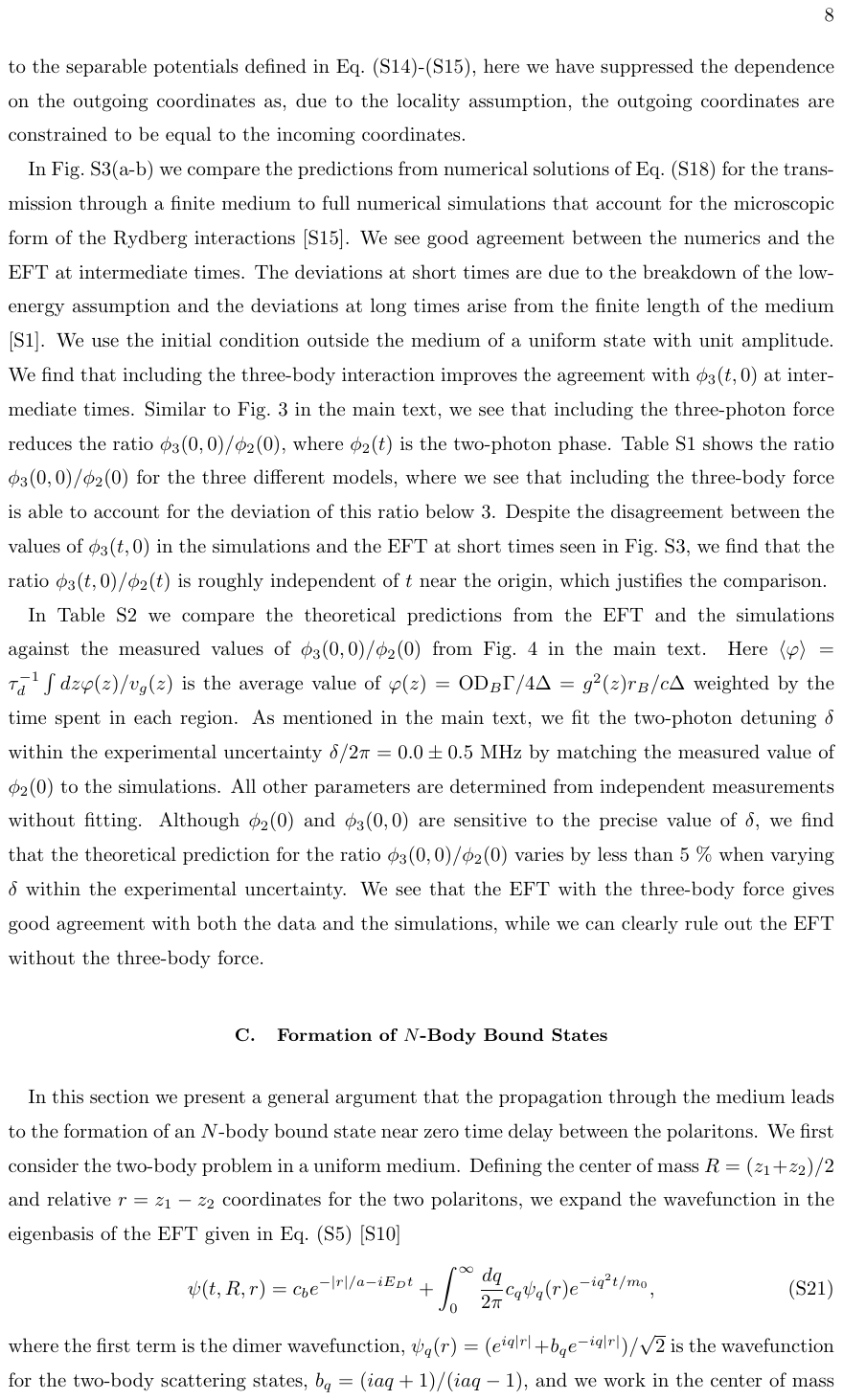}
\end{figure*}

\begin{figure*}
\vspace{-3.35cm}
\hspace*{-2.75cm} 
\includegraphics[page=9]{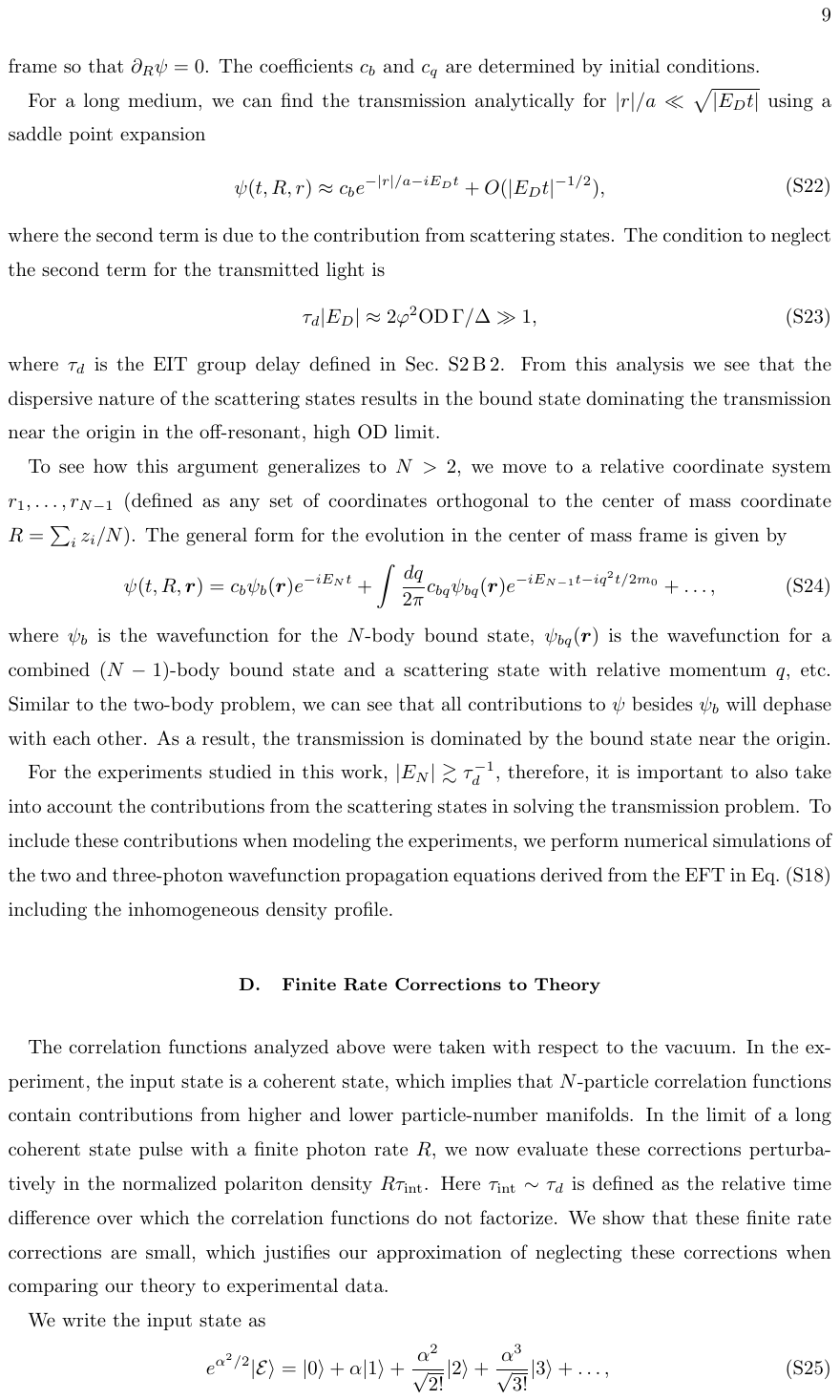}
\end{figure*}

\begin{figure*}
\vspace{-3.35cm}
\hspace*{-2.75cm} 
\includegraphics[page=10]{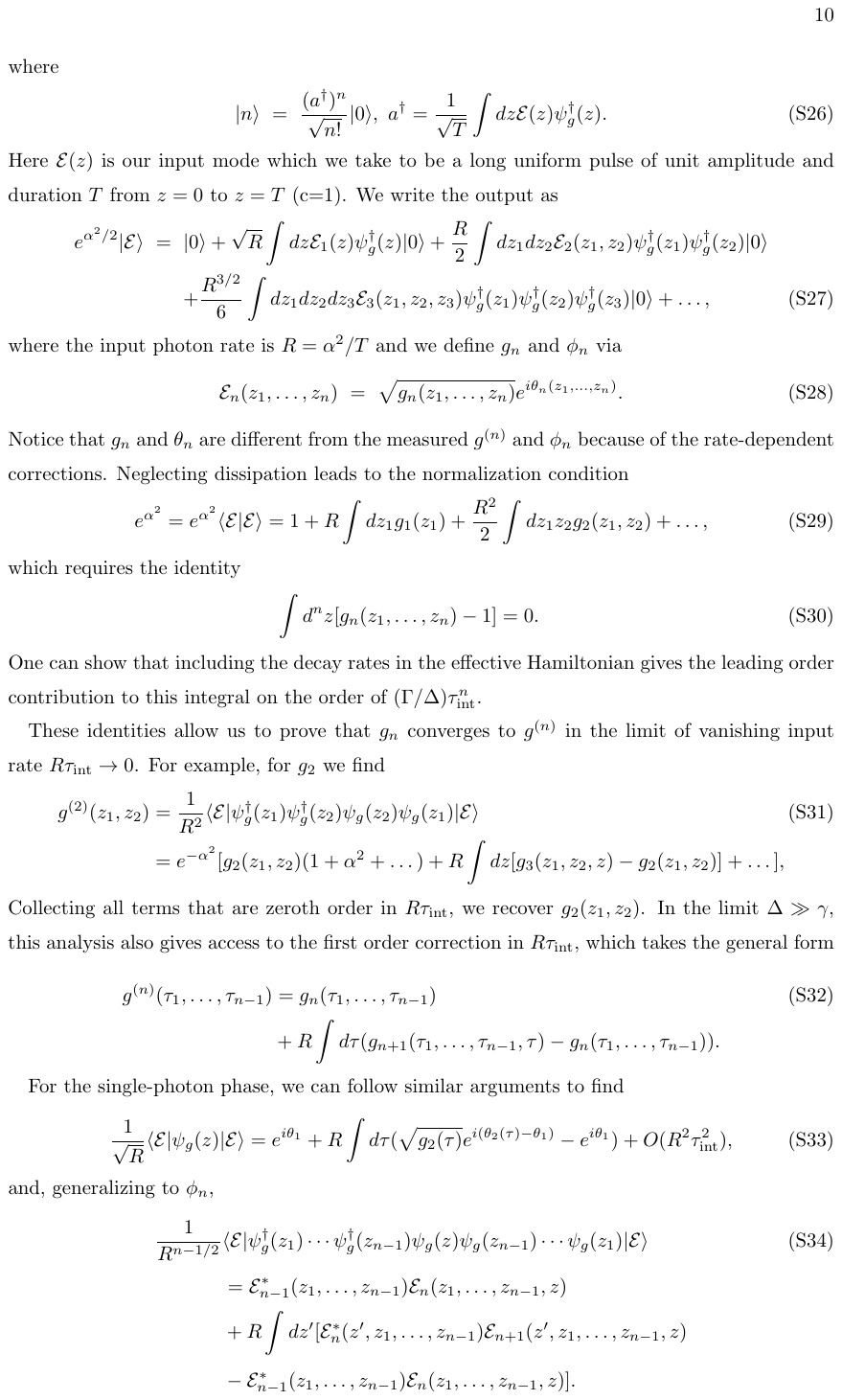}
\end{figure*}

\begin{figure*}
\vspace{-3.35cm}
\hspace*{-2.75cm} 
\includegraphics[page=11]{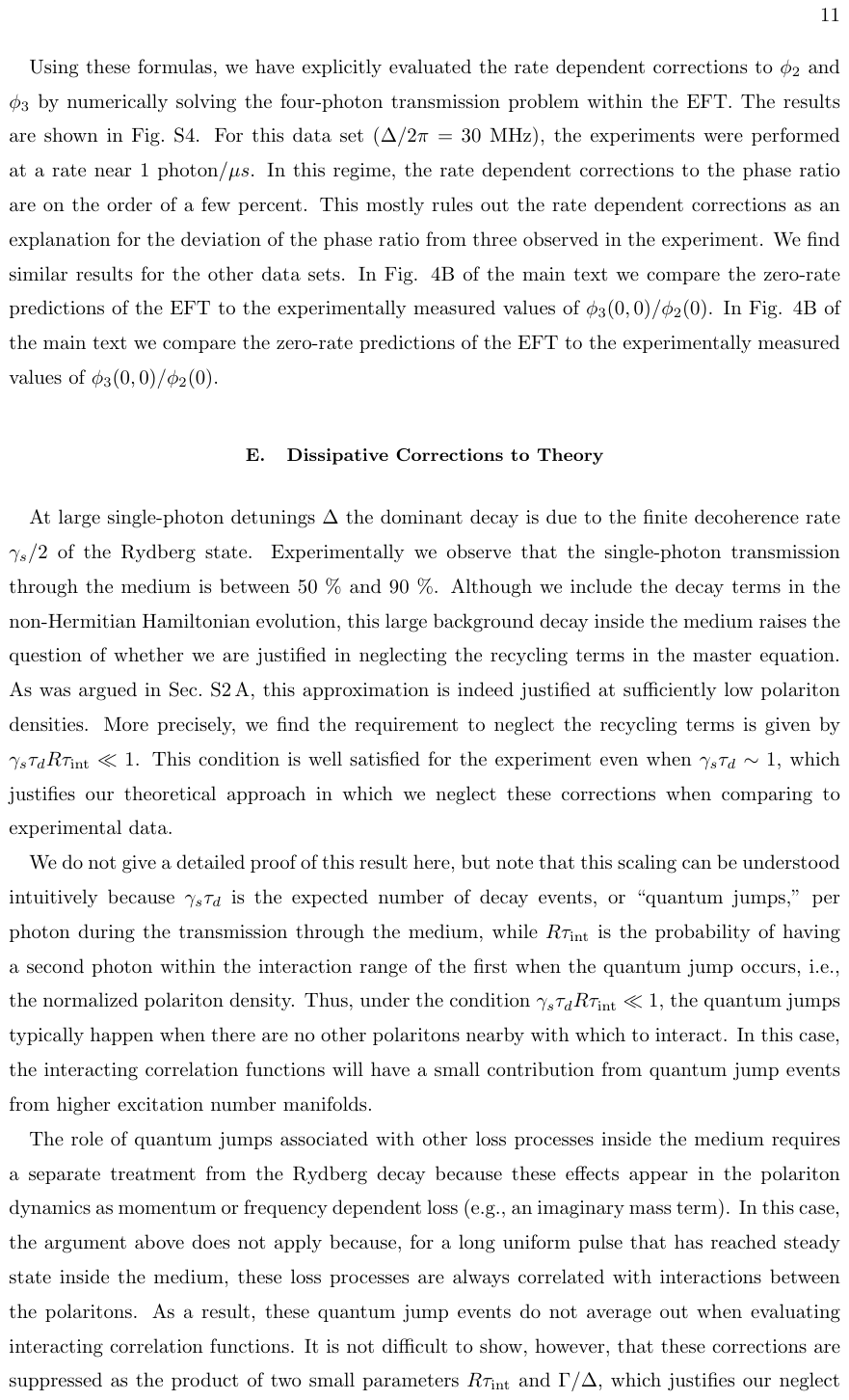}
\end{figure*}

\begin{figure*}
\vspace{-3.35cm}
\hspace*{-2.75cm} 
\includegraphics[page=12]{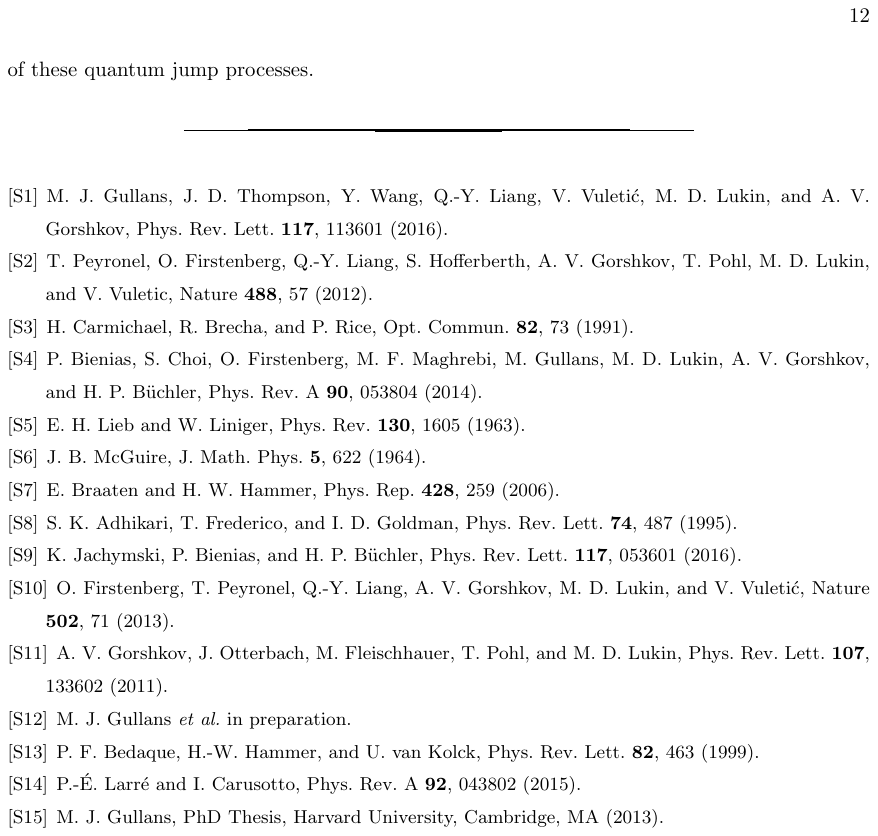}
\end{figure*}

\begin{figure*}
\vspace{-3.35cm}
\hspace*{-2.75cm} 
\includegraphics[page=13]{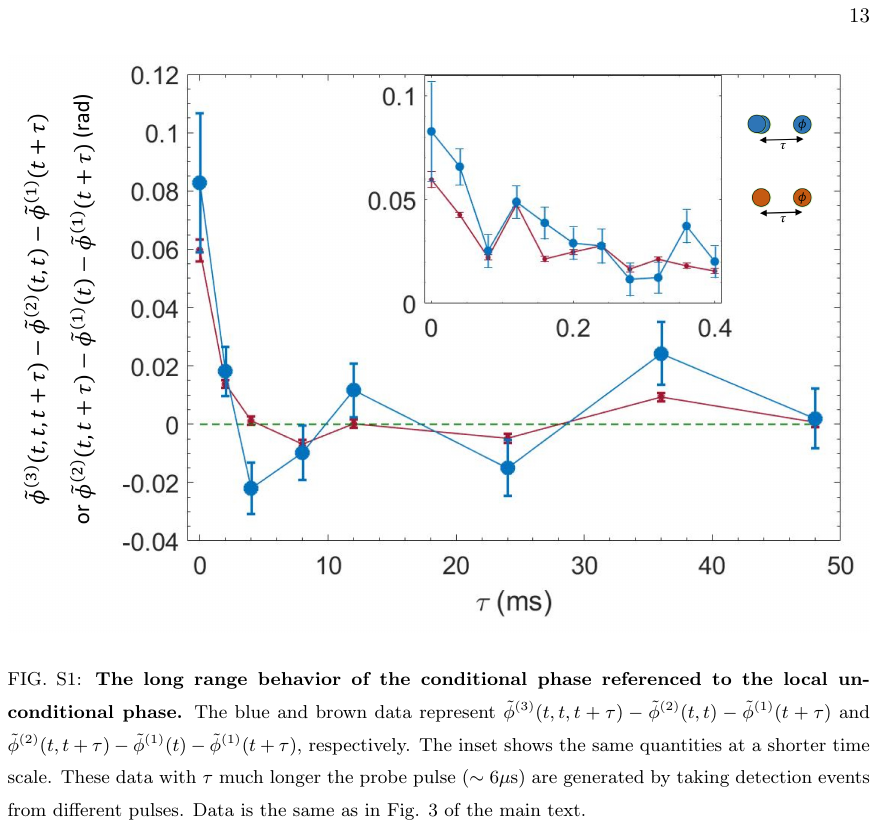}
\end{figure*}

\begin{figure*}
\vspace{-3.35cm}
\hspace*{-2.75cm} 
\includegraphics[page=14]{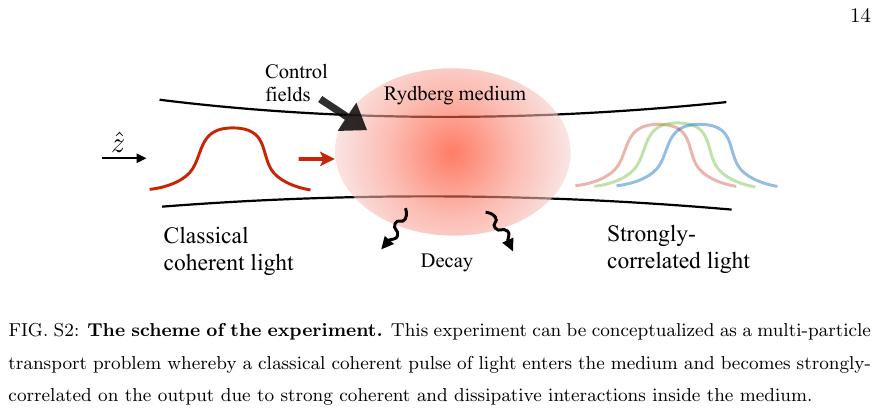}
\end{figure*}

\begin{figure*}
\vspace{-3.35cm}
\hspace*{-2.75cm} 
\includegraphics[page=15]{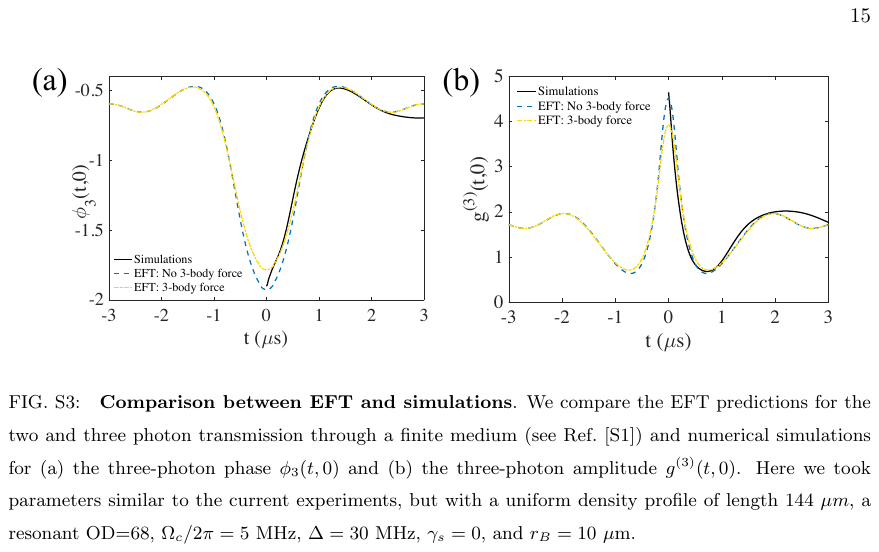}
\end{figure*}

\begin{figure*}
\vspace{-3.35cm}
\hspace*{-2.75cm} 
\includegraphics[page=16]{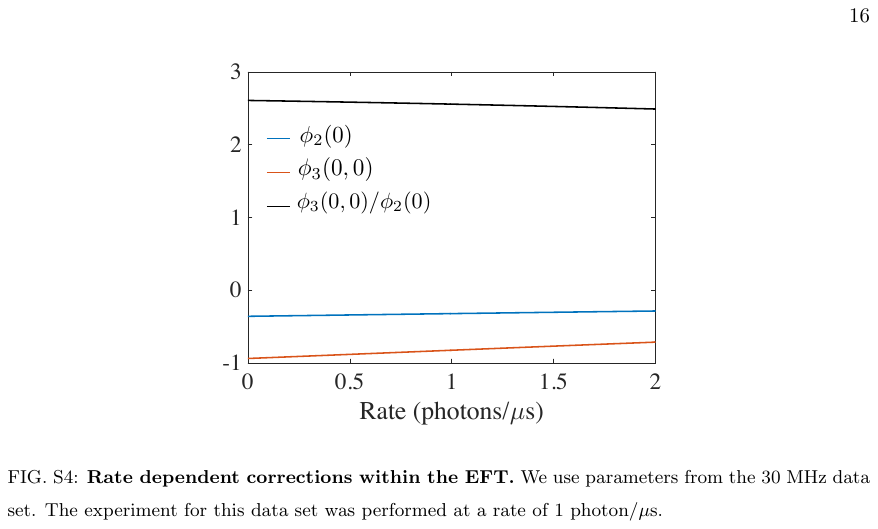}
\end{figure*}

\begin{figure*}
\vspace{-3.35cm}
\hspace*{-2.75cm} 
\includegraphics[page=17]{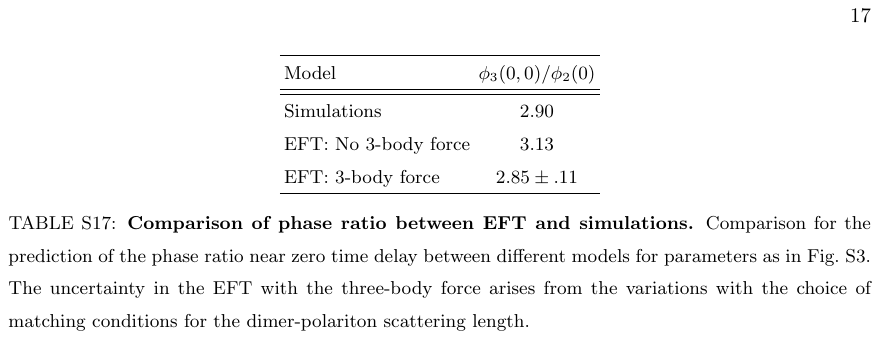}
\end{figure*}

\begin{figure*}
\vspace{-3.35cm}
\hspace*{-2.75cm} 
\includegraphics[page=18]{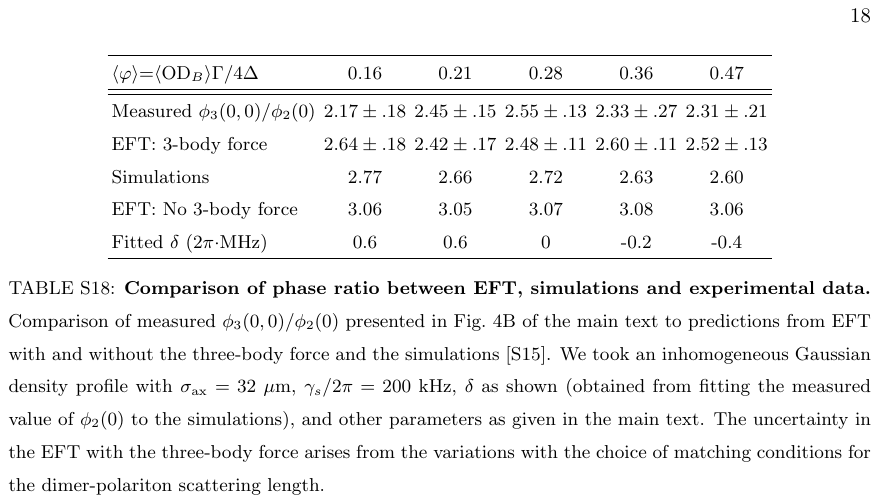}
\end{figure*}

}

\end{document}